# Quantifying macroeconomic expectations in stock markets using *Google Trends*

Johannes Bock

Department of Finance, Warwick Business School, University of Warwick, Scarman Rd, Coventry CV4 7AL, UK, email: j.bock@warwick.ac.uk

May 2018

**Abstract**

Among other macroeconomic indicators, the monthly release of U.S. unemployment rate figures in the Employment Situation report by the U.S. Bureau of Labour Statistics gets a lot of media attention and strongly affects the stock markets. I investigate whether a profitable investment strategy can be constructed by predicting the likely changes in U.S. unemployment before the official news release using *Google query volumes* for related search terms. I find that massive new data sources of human interaction with the Internet not only improves U.S. unemployment rate predictability, but can also enhance market timing of trading strategies when considered jointly with macroeconomic data. My results illustrate the potential of combining extensive behavioural data sets with economic data to anticipate investor expectations and stock market moves.

**Introduction**

Online data potentially offers new opportunities for economists, researchers and investors to investigate and observe shifts in behaviour ahead of more traditional and often backward-looking sources such as consumer surveys or official data. Timely internet-based search data helps to provide "new insights into different stages of large-scale collective decision making" (Preis et al., 2013, p.5) and can be used to analyse people's beliefs, concerns and preferences to ultimately improve forecasts about future behaviour (Ettredge et al., 2005). For example, investigations of *Google Search* data have been applied to predicting exchange rate volatility (Smith, 2012), private consumption (Vosen and Schmidt, 2011) and trading behaviour (Preis et al., 2013). Moreover, Ettredge et al. (2005) were among the first researchers using *Google Search* data to predict macroeconomic indicators including unemployment rates. For the United States, Choi and Varian (2009), and D'Amuri and Marcucci (2017), find that recently laid off people are likely searching for unemployment benefits, vacancies or welfare related topics online and therefore, unemployment forecasts can be improved using *Google* data. Askitas and Zimmermann (2009), Smith (2016), and McLaren (2011) find similar results for Germany and the UK.

As suggested by many researchers, better macroeconomic forecasts could be used to inform economic decision making and compensate for the lagged nature of traditional data sources. Going beyond past studies, this paper seeks to investigate the suggested benefits of incorporating *Google* based macroeconomic forecasts into the investment decision process by testing a hypothetical investment strategy based on monthly U.S. unemployment rate predictions. Since job growth is an important stimulus for the economy and thus, also a driver for investment decisions in financial markets, I expect to be able to construct a profitable trading strategy exploiting a *Google* based forecasting strategy. This study aims to contribute to the literature by shedding light on the impact of macroeconomic news on stock prices. It can be interesting for investors seeking to understand the usefulness of online data for market timing.

**Methods and Results**

In this study, I first analyse the possibility to predict monthly U.S. unemployment rates ($UNEM_t$) using monthly *Google Search query volumes* for a set of 34 search terms and search categories. Monthly and seasonally adjusted U.S. civilian unemployment rate data was obtained for the period from January 2004 until December 2017 from the website of the Federal Reserve Bank of St. Louis[1]. For the same period, using the *Google Trends* service I obtained monthly *Google Web Search volumes* restricted to search requests of users located in the United States[2].

---

[1] U.S. Bureau of Labor Statistics, Civilian Unemployment Rate [UNRATE], retrieved from FRED, Federal Reserve Bank of St. Louis; https://fred.stlouisfed.org/series/UNRATE, March 5, 2018.

A two-step method was used to identify relevant *Google Search* terms for my analysis. First, I identified a base set of 8 search terms by considering past research (D'Amuri and Marcucci, 2017, p.9, Smith, 2016, p.265). Second, using the *GloVe word embeddings*[3] of the root terms "*jobs*", "*unemployment*" and *"unemployed"*, I identified an additional 22 semantically related words by computing the Euclidean distance of 1.9 million English language words to these root terms[4]. I considered word-word co-occurrence based *word embeddings* to be particularly helpful in identifying search terms that are most typically searched along with the root terms or more generally, are related to unemployment. Additionally, following Choi and Varian (2009, p.1), I included four *Google search categories*[5] namely "Jobs" (*60*), "Welfare & unemployment" (*706*), "Work & labour issues" (*703*) and "Job listings" (*960*) in my analysis.

No particular transformation of the data was used, however it is important to note that *Google* data, which is normalized to a range between 0 and 100 with respect to the maximum search volume during a particular time period, is subject to certain limitations and needs to be treated with caution. *Google Trends* data is provided using a sampling method causing it to vary from day to day (Choi and Varian, 2012, p.1) and therefore, there is potentially unobserved heterogeneity in the data.

As suggested in previous research, I will present all my forecasting results using the levels of the monthly U.S. unemployment rate and *Google Search* data (D'Amuri and Marcucci, 2017, p.12-13). Similar to past studies, the $UNEM_t$ forecasting models are based on a linear regression formulation (Eq. 2) which includes a lagged autoregressive component and one lag of the exogeneous variables ($X_{t-1}^i$), which in this study will be the *Google Search volumes* for a particular search term or category $i$ (McLaren, 2011, p.136). As in previous studies, to see whether *Google Trends* data improves prediction accuracy, all models are compared to a *Baseline Model* (Eq. 1), which is a simple AR(1) autoregressive model that includes unemployment level in the previous month as explanatory variable (McLaren, 2011, p.136, Choi and Varian, 2009, p.2).

---

[2] Monthly Google search volumes for 34 search terms and categories, retrieved from Google Trends, https://trends.google.com/trends/, March 5, 2018.
[3] Common Crawl (42B tokens) GloVe word embeddings, retrieved from Stanford University, https://nlp.stanford.edu/projects/glove/, March 4, 2018.
[4] *GloVe word embeddings* are vector representations of words obtained by an unsupervised learning algorithm trained on aggregated global word-word co-occurrence statistics from a text corpus of web data (Pennington et al., 2014).
[5] Using Natural Language Processing *Google Trends* classifies search queries into about 30 broad categories and 250 subcategories (Choi and Varian, 2012).

$$\textit{Baseline Model:} \qquad UNEM_t = \beta_0 + \beta_1 UNEM_{t-1} + \varepsilon_t \qquad (1)$$

$$\textit{Alternative Model:} \qquad UNEM_t = \theta_0 + \theta_1 UNEM_{t-1} + \theta_2^i X_{t-1}^i + \eta_t \qquad (2)$$

It is crucial to note that different from previous research, I did not include the contemporaneous search volumes $X_t^i$ in Eq. 2 due to constraints imposed by my later proposed investment strategy. Therefore, technically speaking this is not a *"nowcasting"* problem like it has been extensively researched in the literature (Choi and Varian, 2009, p.2, Smith, 2016, p.267, D'Amuri and Marcucci, 2017, p.14, McLaren, 2011, p.136). However, as stated by the *U.S. Bureau of Labour Statistics* (*BLS*), unemployment rates ($UNEM_t$) "in month t refer to individuals who do not have a job, but are available for work, in the week including the 12th day of month t and who have looked for a job in the prior 4 weeks ending with the reference week" (D'Amuri and Marcucci, 2017, p.9). Thus, I decided to include one lag of *Google Search volumes* in Eq. 2 to be able to capture the research efforts of individuals looking for a job, who eventually are included in the unemployment rate calculations. In-sample estimation for both models is performed using the full data ranging from January 2004 until December 2017. The out-of-sample testing is conducted using a rolling window procedure, where the models are estimated using the past 36 months to predict $UNEM_t$ in the subsequent month.

The model's forecasting performance and its statistical significance are evaluated by comparing the mean squared errors (MSE) and the squared residuals of each *Alternative Model* to the *Baseline Model* (Fig. 1). According to the Wilcoxon paired signed rank test of squared model residuals, five of the proposed *Alternative Models* achieve statistically significant out-of-sample forecasting improvements, with the largest improvements by including *Google Search volumes* for the search term *laid off* ($laid\ off: W = 5534, p < 0.01; recession: W = 5789, p < 0.01; jobless: W = 5159, p < 0.06; careers: W = 5054, p < 0.1; unemployment: W = 5235, p < 0.05$).

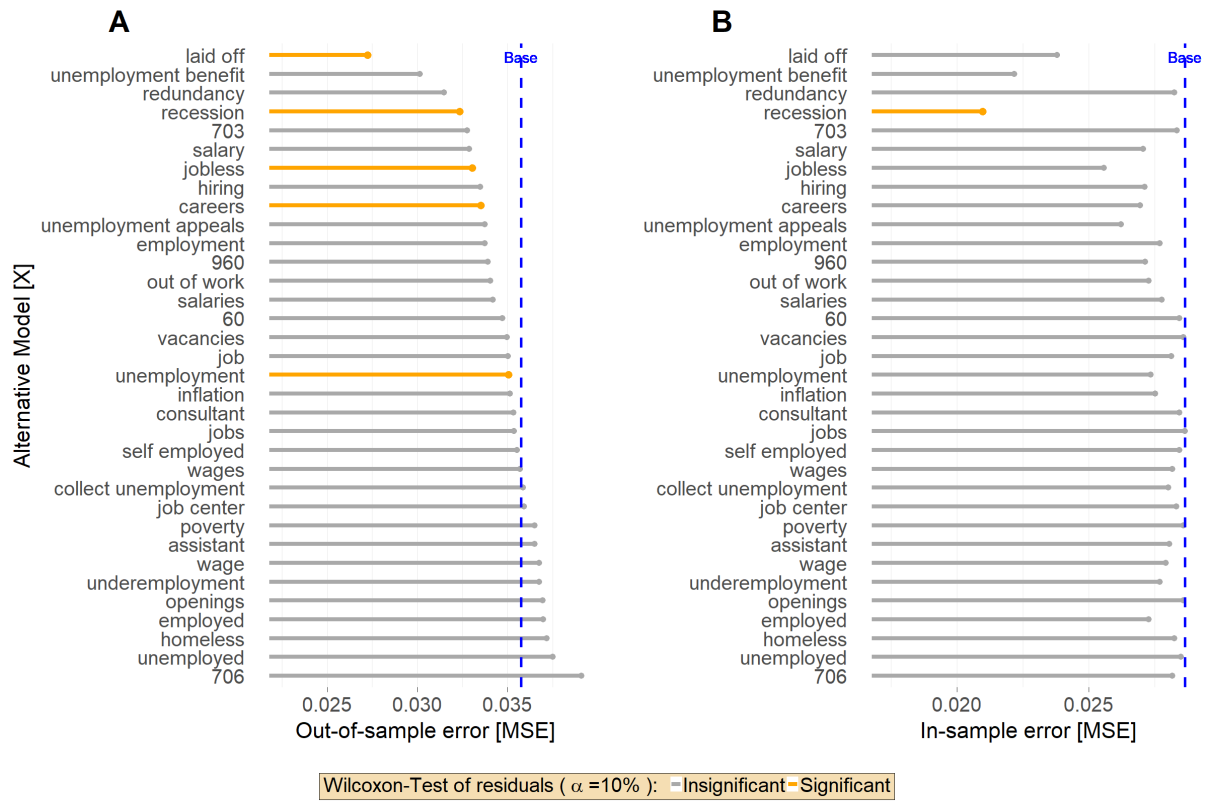

**Figure 1 | Monthly U.S. unemployment rate forecasting improvements using *Google Search* data.** (**A**) In- and (**B**) out-of-sample forecasting error as measured by the mean squared error (MSE) for 34 monthly U.S. unemployment rate ($UNEM_t$) regression models incorporating *Google Trends* data. All 34 models include one autoregressive lag ($UNEM_{t-1}$), and one lag of monthly *Google search* volumes ($X_{t-1}^i$) as explanatory variables for $UNEM_t$. The regressions are estimated in-sample using the full data ranging from January 2004 until December 2017. The out-of-sample testing is conducted using a rolling window procedure, where the models are estimated using the past 36 months to predict $UNEM_t$ in the subsequent month. All models are benchmarked against the baseline autoregressive forecasting model with one lag (vertical blue line). I find that *Google search* data improves MSE against the baseline model both in- and out-of-sample, with statistically significant improvements out-of-sample for five *Google Trends* models ($laid\ off: W = 5534, p < 0.01; recession: W = 5789, p < 0.01; jobless: W = 5159, p < 0.06; careers: W = 5054, p < 0.1; unemployment: W = 5235, p < 0.05; Wilcoxon\ paired\ signed\ rank\ test\ of\ squared\ model\ residuals$).

With the *Google Trends* series, the out-of- sample MSE is decreased by 25.0% and 8.3% against the *Baseline Model*, for the search terms *laid off and jobless,* respectively ($MSE_{Baseline} = 0.036; MSE_{laid\ off} = 0.027; MSE_{jobless} = 0.033$). Additionally, all parameter estimates are statistically significant, even though statistical conclusions about the coefficient estimates need to be treated with wariness, since distributional assumptions do not hold as I will discuss below. My results are largely in line with past research, where D'Amuri and Marcucci (2017, p.16) for instance, find an improvement of 23% over an AR(1) with no *Google* data ($MSE = 0.032$) compared to their best model using *Google* ($MSE = 0.026$). Moreover, note that while some of the *Google Trends* models, such as the *Google Search* category "Welfare & unemployment", increased out-of-sample MSE (Fig. 1A), all of them at least showed equal performance and mostly improved in-sample MSE (Fig. 1B), highlighting the importance of an out-of-sample testing procedure. Overall, I can provide support for the idea that *Google query volumes* for most of the proposed search terms contain predictive information that might be related to job research efforts by individuals who are included in the unemployment statistics, as previously described. Since my regression analysis focuses on unemployment rate and *Google Search* time series data, conclusions about the significance of parameter estimates depend on the stationarity and normality assumption. However, after testing these with an Augmented Dickey-Fuller and Shapiro-Wilk test, respectively, I find that, only for few exceptions, the null hypothesis of a unit root cannot be rejected and normality is strongly rejected. Since *S&P 500* returns are also non-normally distributed (Christoffersen, 2011, p.9), I will only report distribution-free tests in the following analysis.

The second part of my study concerns the implementation of a hypothetical investment strategy that is based on the previously introduced $UNEM_t$ forecasting models (Eq. 2). Therefore, I retrieved daily closing prices $P_d$ of the *S&P 500*, which are adjusted for both dividends and splits, for the period from January 2, 2004 until December 29, 2017[6]. Based on $P_d$, cumulative returns $R_d$ at time $d$ $\prod_{k=1}^{d}[\frac{P_k}{P_{k-1}}] - 1$ were calculated by the cumulative product of daily returns. Moreover, to make the trading strategy operable, I calculated the historical release dates $d_t^*$ of the *Employment Situation report* for month $t$[7].

I propose an investment strategy, where the investment decision for month $t$ is made 15 trading days before the government data release ($d_t^* - 15$) based on the predicted change in $\Delta UNEM_t = \frac{UNEM_t}{UNEM_{t-1}} - 1$. If the unemployment rate is predicted to decrease ($\Delta UNEM_t < 0$), the *S&P 500* is bought at closing price $P_{d_t^*-15-1}$. If however, $\Delta UNEM_t > 0$, I buy if $\Delta UNEM_t - \Delta UNEM_{t-1} < 0$, and short sell otherwise. This second order criteria is based on findings in psychology and behavioural finance that people tend to overweight positive news and underreact to negative news (Easterwood and Nutt, 1999, p.1777). Hence, even though unemployment rates keep rising, if they do not rise as strongly as in the previous period, this is likely to be perceived as positive news and it is interpreted as a slowdown in the crisis, thus I buy. The logic is reversed in upstates of the economy and, therefore, nothing changes in the basic idea that decreasing unemployment is generally good news. In my setup long short and no position if $\Delta UNEM_t = 0$ are allowed and the trading position is held until the next investment decision date. As imposed by the available *Google* data, the one month lagged model formulation and the 36 months of model estimation period, the first trading decision for the March, 2007 government data release on Friday, April 6, 2007 is made at the end of March 15, 2007 and therefore, the investment period starts on March 16, 2007. It is important to note that in my analysis I neglect transaction fees, since the maximum number of transactions per year when using my strategy is only 24. However, such transaction fees would certainly impact profit in a real-world implementation.

---

[6] Adjusted daily closing prices of the *S&P 500*, retrieved from Yahoo! Finance, https://finance.yahoo.com/quote/%5EGSPC/history/, March 5, 2018
[7] *UNEM* release date calculations are based on the fact that data "is typically released on the third Friday after the conclusion of the reference week, i.e., the week which includes the 12th of the month" (also see: https://www.bls.gov/ces/ces_tabl.htm)

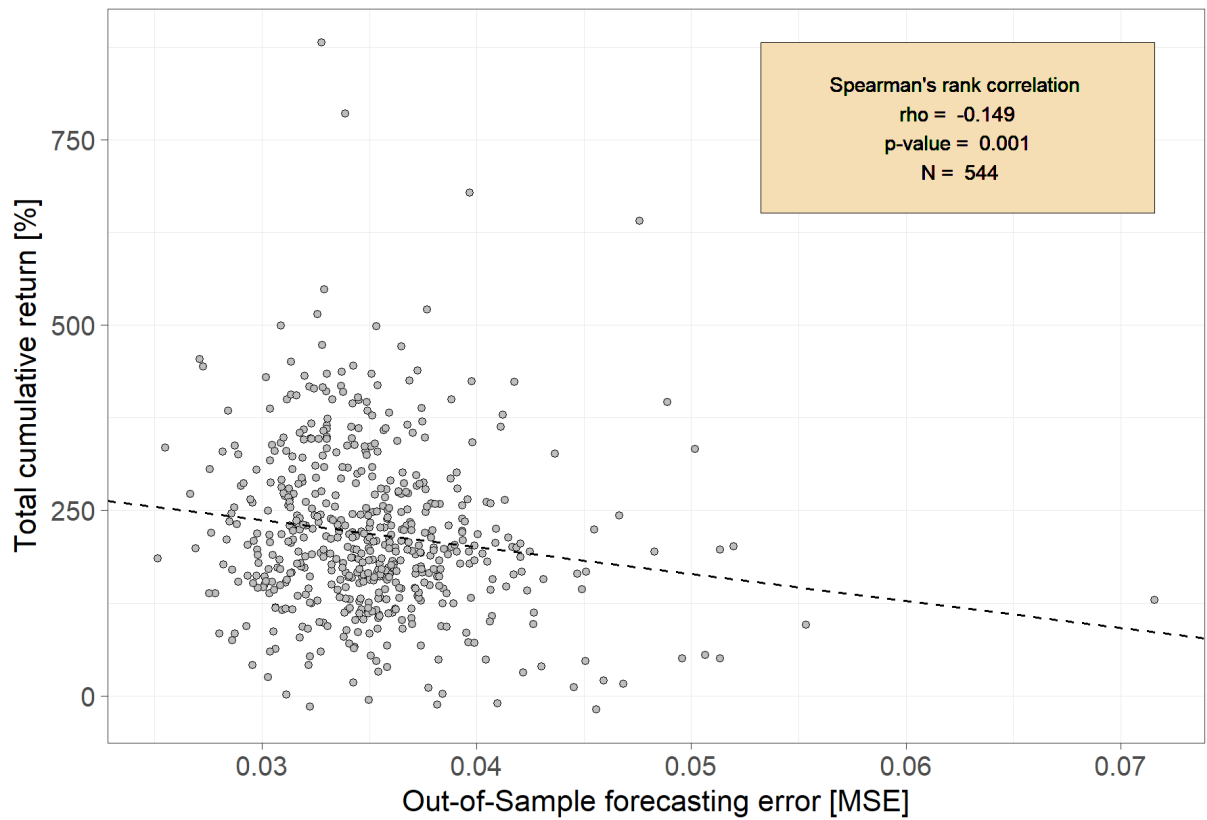

**Figure 2 | Correlation between U.S. unemployment predictability & trading profit.** Correlation between out-of-sample mean squared error (MSE) in monthly U.S. unemployment rate ($UNEM_t$) predictions and cumulative trading returns. Returns are obtained by investment strategies which trade 15 trading days before the government data release and are based on 544 different $UNEM_t$ forecasting models incorporating *Google Trends data* from 34 search terms/categories. The prediction model specifications vary with respect to their number of autoregressive lags ($UNEM_{t-k}, k \in [1,2]$), *Google search volume* lags ($X^i_{t-k}, k \in [1,2]$) and out-of-sample estimation window size ($m \in [12, 24, 36, 48]$). I find a significant negative correlation between forecasting error and trading returns, suggesting that the better U.S. unemployment predictability, the larger trading profits can be made ($\rho = -0.15, \ p < 0.01, \ Spearman's \ rank \ correlation$).

The trading strategy was implemented based on 544 variations of the $UNEM_t$ forecasting models. Differently from Eq. 2, these models varied with respect to their number of autoregressive lags ($UNEM_{t-k}, k \in [1,2]$), *Google search volume* lags ($X^i_{t-k}, k \in [1,2]$) and out-of-sample estimation window size ($m \in [12, 24, 36, 48]$). As shown in Fig. 2, I find that the correlation between out-of-sample MSE and cumulative trading returns over the whole investment period is significantly negative ($\rho = -0.15, \ p < 0.01, \text{Spearman's rank correlation}$). This suggests that the better U.S. unemployment rate predictability, the larger trading profits can be made.

Moreover, the performance of the investment strategy is evaluated on the basis of the difference in overall returns against four different benchmarks (Fig. 3). First, it is compared to a strategy that is identical but based on forecasts of the *Baseline Model*. Second, I contrast a strategy that buys or sells at closing price on the day of the data release, based on the actual data.

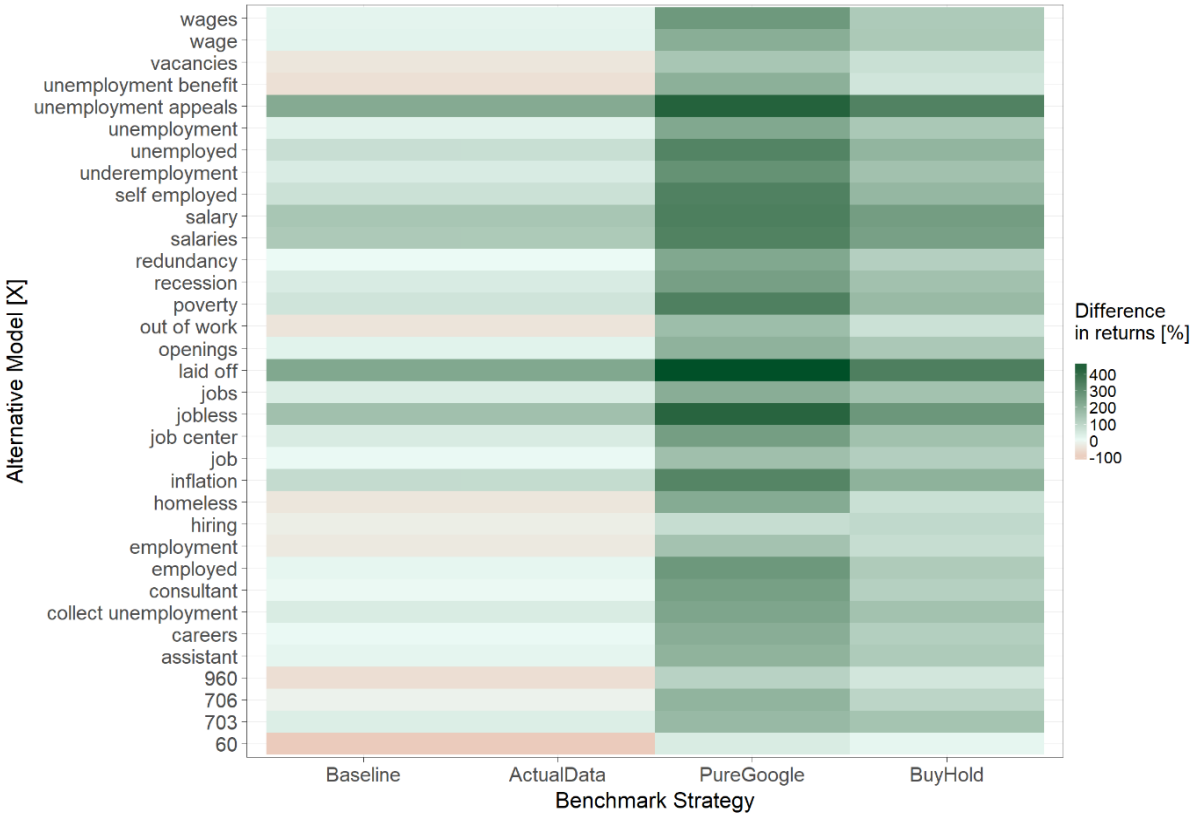

**Figure 3 | Performances of *Google Trends* investment strategies against their benchmarks.** Difference in cumulative returns between 34 investment strategies based on U.S. unemployment rate ($UNEM_t$) forecasting models and their benchmark trading strategies. The 34 investment strategies are based on prediction models that include one autoregressive lag ($UNEM_{t-1}$), and one lag of monthly *Google search volumes* ($X_{t-1}^i$) as explanatory variables for $UNEM_t$ and trade 15 trading days before the government data release. The benchmarks include a strategy based on the baseline autoregressive forecasting model with one lag (left), a strategy based on the actual changes in monthly $UNEM_t$, a strategy based purely on the *Google search volumes* of the respective search term/category and a "buy and hold" strategy (right). The difference in cumulative returns is calculated using the respective overall portfolio returns over the entire investment period of my study from March 16, 2007 until December 29, 2017. I find that returns from the *Google Trends* strategies are significantly higher overall than returns from the benchmark strategies ($<R>_{Baseline} = 2.11; W = 414, p < 0.03; <R>_{ActualData} = 2.11; W = 413, p < 0.03; <R>_{BuyHold} = 0.92; W = 595, p < 0.01, one-sample\ Wilcoxon\ signed-rank\ test;$ and $W_{PureGoogle} = 595, p < 0.01, two\text{-}sample\ Wilcoxon\ paired\ signed\ rank\ test$).

Third, I include a strategy, which is purely based on the change in *Google Search volumes* for the respective search term, since similar strategies were proposed in past research (Preis et al., 2013). In this case I sell at the beginning of month $t$ if search volumes have increased in month $t-1$ and vice versa for the long position. Finally, the "buy and hold strategy" will serve as the benchmark for the market performance. I find that returns from the proposed *Google Trends* strategies are significantly higher than returns from the benchmarks ($<R>_{Baseline} = 2.11; W = 414, p < 0.03; <R>_{ActualData} = 2.11; W = 413, p < 0.03; <R>_{BuyHold} = 0.92; W = 595, p < 0.01, one-sample\ Wilcoxon\ signed-rank\ test;$ and $W_{PureGoogle} = 595, p < 0.01, two\text{-}sample\ Wilcoxon\ paired\ signed\ rank\ test$). However, my results show that performance of the *Google Trends* strategy differs with the search term chosen. In Fig. 3 I investigate these differences and conclude that, even though there is some variation, when compared to the "buy and hold" and purely *Google Trends* based strategy all 34 *Alternative Models* yield higher overall returns of up to 500% (*laid off*). But, the *Baseline Model* and the actual data strategy are more competitive. To beat their performance, more accurate $UNEM_t$ prediction models are required and therefore, the *laid off* and *jobless* models, which significantly improved out-of-sample MSE (see Fig. 1A), can compete, for instance.

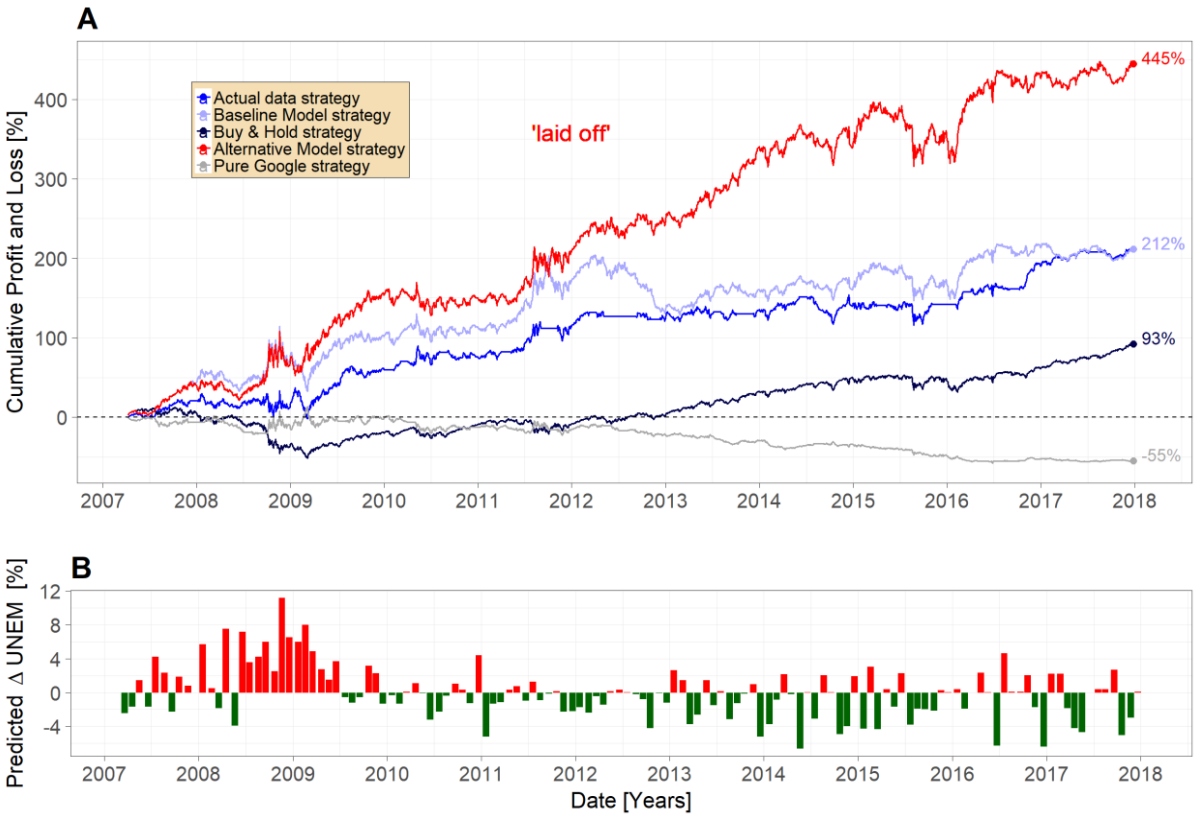

**Figure 4 | Performance of an investment strategy based on unemployment predictions using *Google Trends* data.** (**A**) Profit and loss for an investment strategy based on (**B**) the predicted month-over-month change in U.S. unemployment rates ($\Delta UNEM_t$) plotted as a function of time. $UNEM_t$ predictions are based on *Google search volumes* for the search term *laid off*, the keyword, which in my analysis gave the largest increase in out-of-sample forecasting accuracy (Fig. 1A). Investment decisions are made 15 trading days before the government data release and are based on prediction models that include one autoregressive lag ($UNEM_{t-1}$), and one lag of *monthly Google search volumes* ($X_{t-1}^{laid\,off}$) as explanatory variables for $UNEM_t$. This is compared to a strategy based purely on the *Google search volumes* for *laid off* (grey line), a strategy based on the actual changes in monthly $UNEM_t$ (dark blue line), a strategy based on the baseline autoregressive forecasting model with one lag (light blue line) and a "buy and hold" strategy (black line). The *Google Trends* strategy using the search volume of the term *laid off* would have yielded a profit of 445% (red line).

Some of the models, such as the one based on the search category "jobs", would have yielded about 108% lower returns than the *Baseline* or actual data strategy. Fig. 4 shows the cumulative return as a function of time for the search term *laid off*, which in my analysis gave the largest increase in out-of-sample forecasting accuracy (Fig. 1A). The use of this *Google Trends* strategy would have increased the value of a portfolio by 445%. Note that in hindsight all active trading strategies, including all benchmarks except the "buy and hold" strategy, would have avoided large losses during the financial crisis 2008-2010. However, only the *Alternative Model* strategy continued to benefit from rising stock markets, even though it could also not prevent losses during 2015 and beginning of 2016.

**Discussion**

Monthly government data releases on economic conditions are crucial information for investors, which drive investment decisions in financial markets, but unfortunately are lagged by several days and not readily available. Data on human interaction with *Google Search* has been shown to correlate with the current level of economic activity and predict certain economic indicators such as unemployment rates. Therefore, I have investigated the possibility of exploiting this information in the present study by suggesting a hypothetical investment strategy based on monthly U.S. unemployment rate forecasts.

I have estimated and tested 34 forecasting models using *Google query volumes* for search terms and categories related to employment and have found that unemployment predictions can significantly be improved using *Google* data. Additionally, I have demonstrated that my forecasting models could have been used in the construction of profitable trading strategies. I have shown that portfolio returns of the proposed strategies not only significantly correlated with unemployment forecasting accuracy, but also significantly outperform several benchmark strategies including a "buy and hold" strategy, and strategies purely based on actual unemployment or *Google* data. Hence, I find support for the idea that trading behaviour in financial markets can be anticipated by forecasting changes in important economic indicators.

I offer one possible interpretation of my results within the context of investor expectations. Investors closely monitor job growth since it is an important stimulus for economic growth and a general driver of consumer confidence and spending in the economy (Ludvigson, 2004, p.31). Therefore, it is no surprise that changes in unemployment statistics are affecting stock markets. However, since by definition a profitable investment strategy needs to anticipate investor behaviour – or, more precisely, investor expectations -, my findings suggest that expectations of unemployment levels might be already formed much before the release of new Employment Situation reports. Therefore, I suggest that *Google search volumes* are a good measure of latent factors that ultimately impact investor expectations, and when jointly considered with unemployment data this results in a good approximation of investor expectations of unemployment levels. This is line with McLaren (2011), who argues that *Google Trends* data may contain information above and beyond those provided by survey indicators. However, it is also important to note that *Google Search* behaviour is constantly changing and issues like herd behaviour can make initially observed patterns obsolete. Such issues have considerable implications for forecasting ability and for example, have been suggested as a key reason behind the persistent errors by *Google Search* data in flu predictions in recent years (Ormerod et al., 2014).

To conclude, my results illustrate the potential of combining extensive behavioural data sets with economic data to anticipate investor expectations and stock market moves, but this should also be treated with wariness due to fluctuations in *Google Search* behaviour. Future research could extend my analysis to other important economic statistics such as inflation or retail sales and investigate if investor expectations can be anticipated more widely.